\documentclass[12pt]{iopart}

\usepackage{iopams}

\usepackage{epsfig}
\usepackage{psfrag}
\usepackage{graphicx}
\usepackage{cite}

\newcommand{\half}{\frac{1}{2}}

\begin{document}

\title{$c=1$ Boundary Conformal Field Theory Revisited}

\author{Kristj\'an R Kristj\'ansson and L\'arus Thorlacius}
\address{University of Iceland, Science Institute,
Dunhaga 3, 107 Reykjavik, Iceland}
\eads{kristk@hi.is, lth@hi.is}

\begin{abstract}
Correlation functions of discrete primary fields in the $c=1$ 
boundary conformal field theory of a scalar field in a critical
periodic boundary potential are computed using the underlying 
$SU(2)$ symmetry of the model.  Bulk amplitudes are unambigously 
determined and we give a prescription for amplitudes involving 
discrete boundary fields.
\end{abstract}

\section{Introduction}
\label{introduction}

We consider a two-dimensional scalar field $\Phi(z,\bar z)$
defined on the upper-half-plane, $\mathrm{Im}\, z>0$, which is 
free in the two-dimensional bulk but subject to a periodic
boundary potential. The action is
\begin{equation}
  \label{action}
  S = \frac{1}{4\pi} \int \rmd^2 z\: \partial \Phi \bar \partial \Phi
    - \frac{1}{2} \int \rmd\tau \left( g\, \rme^{\rmi\Phi(0)/\sqrt{2}} 
                  + \bar g\, \rme^{-\rmi\Phi(0)/\sqrt{2}}\right)
\end{equation}
where $g$ is a complex parameter which dials the strength and the phase 
of the periodic boundary interaction.  The period of the potential is 
chosen such that the interaction has dimension one under boundary scaling.
This ensures that the boundary theory preserves half of the conformal
symmetry of the bulk system.  
 
This model arises in various contexts, for example, in connection 
with critical behavior in dissipative quantum 
mechanics \cite{Caldeira:1983uj,Fisher:1985,Guinea:1985},
quantum Hall edge states \cite{Kane:1992}, and open string theory 
in an on-shell tachyon background \cite{Callan:1990sf}.
More recently a $c =1$ boundary conformal field theory involving a scalar 
field with a `wrong-sign' kinetic term and an exponential boundary 
interaction has been applied to the rolling tachyon field of an S-brane 
or an unstable 
D-brane \cite{Sen:2002nu,Strominger:2002pc,Gutperle:2003xf,Larsen:2002wc}.
Such a theory is related by analytic continuation of the field to the
boundary theory considered here.

~A conformal field theory analysis of the model (\ref{action}), 
carried out a decade ago \cite{Callan:1994mw,Callan:1994ub,Polchinski:1994my}, 
revealed an underlying $SU(2)$ symmetry which 
allows many exact results to be established, including exact boundary
states and the one-loop partition function.  In this lecture we report 
on ongoing work aimed at calculating correlation functions of both bulk 
and boundary fields. The general problem involving fields carrying 
arbitrary momenta remains unsolved but we give a prescription for 
correlation functions of an important sub-class of fields, the so-called
discrete primaries.  

\section{Left-moving current algebra}
\label{currentalgebra}

We take the scalar 
field to be non-compact, {\it i.e.}\ taking value in $\mathbb{R}$, although 
the compact case, where the field takes value on a circle of radius $R$, 
is also of interest.  
In the absence of the boundary interaction, the field $\Phi(z,\bar z)$
satisfies a Neumann boundary condition at $\mathrm{Im}\, z=0$, and this
is conveniently dealt with using the so-called doubling trick or method 
of images (see f.ex.\ \cite{DiFrancesco:1997nk,Polchinski:1998rq}). 
The theory on the upper half-plane is then mapped into a chiral
theory on the full complex plane and the boundary eliminated.
To see how this works we note that away from the boundary the free 
field can be written as a sum of left- and right-moving components,
$\Phi(z,\bar z) = \phi(z) + \bar \phi(\bar z)$.
For $g=0$ the Neumann boundary condition on the real axis 
$\bar \phi(\bar z) - \phi(z)\vert_{z=\bar z}=0$ 
determines the right-moving field in terms
of the left-moving one. We can therefore reflect the right-moving field 
through the boundary and represent it as a left-moving field in the 
lower-half-plane.  The theory then contains only left-moving fields, but
a left-moving field at $z^*$ in the unphysical lower-half-plane is to be 
interpreted as a right-moving field at $z$ in the physical 
region.\footnote{Our notation follows that of \cite{DiFrancesco:1997nk}.
Both $\bar z$ and $z^*$ denote the complex conjugate of $z$.  We use
$\bar z$ for the argument of a right-moving field 
$\bar\phi(\bar z)$, but $z^*$ for that of a left-moving image field 
$\phi(z^*)$.}

More generally, any quasi-primary field in the bulk separates into 
left- and right-moving parts
$\Psi_{h,\bar h}(z,\bar z) = \psi_h(z)\bar \psi_{\bar h}(\bar z)$.
Using the doubling trick 
$\bar \psi_{\bar h}(\bar z)$ becomes a left-moving field 
$\psi_{\bar h}(z^*)$, with holomorphic dimension $\bar h$.
A bulk $n$-point function 
$\langle \Psi_{h_1,\bar h_1}(z_1,\bar z_1) \cdots
\Psi_{h_n,\bar h_n}(z_n,\bar z_n)\rangle$ 
in the original theory on the upper-half-plane then becomes a
$2n$-point function of holomorphic fields
$\langle \psi_{h_1}(z_1)\psi_{\bar h_1}(z_1^*) \cdots
\psi_{h_n}(z_n)\psi_{\bar h_n}(z_n^*)\rangle$ 
on the infinite plane.

It turns out that the doubling trick can be applied even when
the boundary interaction in (\ref{action}) is turned on.  This is 
because the boundary potential can in fact be expressed in terms of 
the left-moving field alone
\begin{equation}
\label{bpotential}
-\frac{1}{2} 
\left( g\, \rme^{\rmi\sqrt{2}\phi(z)} 
+ \bar g\, \rme^{-\rmi\sqrt{2}\phi(z)}\right)
\bigg\vert_{\mathrm{Im}(z) = 0} \>.
\end{equation}
The operators appearing in the interaction are 
currents of a left-moving $SU(2)$ algebra
\begin{equation}
J_\pm = \rme^{\pm \rmi\sqrt{2}\phi(z)}, \qquad 
J_3 = \rmi\,\partial \phi(z)/\sqrt{2}.
\end{equation}
In the boundary action (\ref{action}) the $J_+$ and $J_-$ currents are 
integrated along the real axis and in a perturbative 
expansion of a correlation function such integrals are repeatedly 
inserted into the amplitude.   As usual, divergences arise when operator 
insertions coincide but, by a clever choice of regularization, Callan 
\etal~\cite{Callan:1994ub} were able to sum the perturbation
series explicitly to obtain the exact interacting boundary state.  It
turned out to be remarkably simple, with the net effect of the interaction 
being a global $SU(2)$ rotation, 
$U(g) = \exp \pi \rmi (g J_+ + \bar g J_-)$,
acting on the free Neumann boundary state.

\section{Bulk primary fields}
\label{bulkprimaries}

The bulk theory is that of a free boson.  It contains holomorphic primary 
fields $\rme^{\rmi k\phi(z)}$ with conformal weight $h=k^2/2$ for all 
$k\in \mathbb{R}$ and also the corresponding anti-holomorphic fields.  
At special values of the momentum, $k = \sqrt{2}\,j$, where $j$ is an 
integer or integer-plus-half, some descendant states have vanishing norm 
and new primary fields appear, the so-called discrete primaries 
\cite{Kac:1979}, which come in $SU(2)$ multiplets labelled by $j$ and $m$, 
with $-j \le m \le j$.  The discrete fields $\psi_{j m}(z)$ in a given 
$SU(2)$ multiplet are degenerate in that they all have conformal weight 
$h=j^2$.  They are composite fields made from polynomials in 
$\partial\phi$, $\partial^2\phi$, {\it etc.}\ accompanied by 
$\rme^{\rmi\sqrt{2}m\phi}$, and normal ordered with respect to the 
free holomorphic propagator 
$\langle\phi(z)\phi(z')\rangle = - \log (z-z')$.

The discrete fields have the following representation \cite{Klebanov:1991hx},
\begin{equation}
\psi_{jm}(z) \sim
\left(\oint {\rmd w\over 2\pi \rmi} \rme^{-\rmi\sqrt{2}\phi(w)}\right)^{j-m}
\rme^{\rmi\sqrt{2}j\phi(z)} \,,
\end{equation}
where the lowering current is integrated along nested contours 
surrounding $z$. 
A discrete bulk primary is constructed from a pair of 
holomorphic and anti-holomorphic primaries,
\begin{equation}
\label{discretebulkfield}
\Psi_{j \bar\jmath m}(z,\bar z) =
\psi_{j m}(z)\bar\psi_{\bar\jmath m}(\bar z) \,.  
\end{equation}
A priori, the left- and right-moving fields can carry different $SU(2)$ 
labels.  However, the spin $h-\bar h=j^2-\bar\jmath^2$ takes an unphysical
value unless $j-\bar\jmath\in \mathbb{Z}$.  
We are considering a non-compact free boson, which has no winding states, 
so we must also require $p_L=p_R$, which amounts to $\bar m=m$.  
Finally we have $-j_0\le m \le j_0$ where $j_0=\min{(j,\bar\jmath)}$.

\section{Boundary fields}
\label{boundaryfields}

When a bulk field approaches the boundary at $z=\bar z$, new divergences
appear that are not removed by the bulk normal ordering.  This is a 
general feature of conformal field theories with boundaries and signals
the presence of so called boundary operators.
The boundary conditions to either side of a boundary
operator can be different, in which case it is called a boundary 
condition changing operator (see for example \cite{DiFrancesco:1997nk}).  
In a general boundary conformal field theory a bulk field approaching the 
boundary can be expanded in terms of boundary fields \cite{Diehl:1981},
\begin{equation}
\label{bbope}
\Psi_{h,\bar h}(z, \bar z) =
\sum_i \frac{A^i_{h,\bar h}}{(z-\bar z)^{h + \bar h-\Delta_i}}\, \Psi^B_i(x)\,,
\end{equation}
where $x = \half (z+\bar z)$.  The $\Psi^B_i(x)$ are boundary fields (possibly
boundary condition changing) with boundary scaling dimensions $\Delta_i$ and the 
$A^i_{h,\bar h}$ are called bulk-to-boundary operator product coefficients.  

In addition to the bulk-to-boundary OPE, the boundary fields form an 
operator product algebra amongst themselves,
\begin{equation}
\label{boundaryope}
\Psi^B_i(x) \, \Psi^B_j(x') = \sum_k 
\frac{C_{ijk}}{(x-x')^{\Delta_i+\Delta_j-\Delta_k}} \,\Psi^B_k(x')\,.
\end{equation}
The boundary OPE coefficients $C_{ijk}$ and the bulk-to-boundary
OPE coefficients $A^i_{h,\bar h}$, along with the boundary scaling
dimensions $\Delta_i$, are characteristic data of a given boundary
conformal field theory.
In particular, the boundary scaling dimension $\Delta_i$ of a boundary 
operator $\Psi^B_i(x)$ is given by the energy eigenvalue of the 
corresponding open string state, where the open-string Hamiltonian is 
the $L_0$ generator of the Virasoro algebra that is preserved by the 
conformally invariant boundary conditions.  

\begin{figure}
\begin{center}
\psfrag{En}{\footnotesize $\Delta$}
\psfrag{k}{\footnotesize $k$}
\psfrag{1}{\footnotesize 1}
\psfrag{4}{\footnotesize 4}
\includegraphics[width=\linewidth]{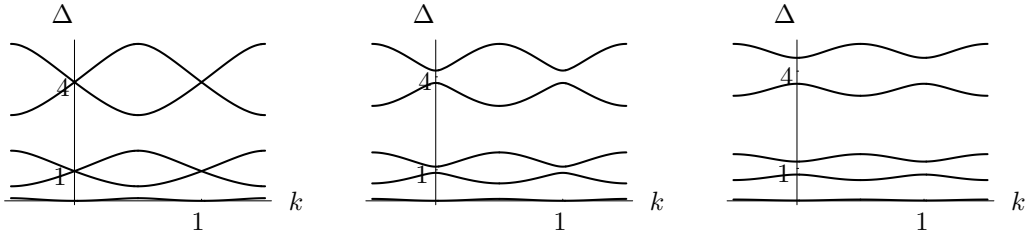}
\end{center}
\caption{
The first few bands of the open string spectrum for different values 
of the coupling constants. One is held fixed at $g'=0.2$ but the other
one takes the values a)~$g=0.2$, b)~$g=0.3$, c)~$g=0.4$.
\label{fig:genspectrum}}
\end{figure}

In the theory at hand, boundary conditions are labelled by the boundary 
coupling $g$ in (\ref{action}).  A boundary condition changing operator
that changes $g$ to $g'$ at the insertion point 
corresponds to an open string state where the two string endpoints 
interact with boundary potentials of different strength $g$ and $g'$.
To work out the open string spectrum we find it convenient to re-express 
the interacting boundary
theory in terms of free fermions as shown in \cite{Polchinski:1994my}.  
The $SU(2)$ currents are bi-linear in the fermions and the boundary 
interaction may be viewed as a localized mass term.  The open string 
spectrum is then found by solving a straightforward eigenvalue problem 
for the fermions.  We will not repeat the construction here but simply 
quote the result.  In \cite{Polchinski:1994my} both string endpoints 
were taken to interact with the same boundary potential, {\it i.e.}\ it 
was assumed that $g=g'$.  In this case the partition function may be 
written
\begin{equation}
\mathcal{Z} = {\sqrt{2}\over \eta(q)} \int_{-1/2}^{1/2} \frac{dk}{2\pi}
      \sum_{m=-\infty}^{\infty} q^{(\lambda + m)^2} \,,
      \end{equation}
where $\eta(q)$ is the Dedekind eta function, $q=e^{-\pi\beta/\ell}$
with $\ell$ the parameter length of the open string, and $\lambda$ is 
related to the target space momentum $p=\sqrt{2}k$ by
\begin{equation}
\label{ggspectrum}
\sin \pi \lambda = \cos \pi |g| \sin \pi k  \,.
\end{equation}
The value of $\lambda$ is determined by continuity from $\lambda =k$ at $g=0$.
The scaling dimensions of the corresponding boundary operators are given by
$\Delta=\left(\lambda(k) + m\right)^2$ with $m\in \mathbb{Z}$.
The spectrum obtained from (\ref{ggspectrum}) is shown in the left-most 
graph in \fref{fig:genspectrum}.
The free spectrum, $\Delta = \half p^2$, has split into bands with forbidden 
gaps in energy in between them.  
Note that the energy eigenvalues are invariant under $k \to k + n$ for any 
integer $n$. 
The boundary potential breaks 
translation invariance in the target space to a discrete subgroup and 
target space momentum is only conserved 
mod $\sqrt{2}\,\mathbb{Z}$ in our units in the interacting system. 
The momentum of a given operator can therefore always be shifted
into the so-called first
Brillouin zone, $-\half\leq k \leq \half$.  The spectra in
\fref{fig:genspectrum} are displayed in an extended zone scheme,
with the periodicity appearing explicitly.

It is interesting to note that bands also appear in the open string
channel of the boundary conformal field theory of a free boson 
compactified on a circle, when the radius of the circle is an irrational
multiple of the self-dual radius $R_{\rm sd}=\sqrt{2}$
\cite{Friedan:1999,Janik:2001hb}. Like the system we are considering 
here, those theories admit a one-parameter family of boundary states that 
interpolate between Neumann and Dirichlet boundary conditions, but it is 
unclear to us at present how deep the parallels between the systems run.

The band spectrum described by (\ref{ggspectrum}) is
a special case of a more general structure that appears when we allow for 
open strings with different boundary coupling, $g$ and $g'$, at the two
endpoints. For simplicity we take $g,g' \in \mathbb{R}$, with $g\geq g'$.
The fermion eigenvalue problem solved in \cite{Polchinski:1994my} can 
easily be extended to cover this case also.  The only modification is to 
change $g$ to $g'$ in one of the boundary mass terms for the worldsheet
fermions in equation (29) of that paper, leaving the other one 
unchanged.  The eigenvalue equation for the spectrum then becomes
\begin{equation}
\label{eq:generalspectrum}
\sin^2 \pi\lambda = \sin^2 \pi g_- \cos^2 \pi k 
                  + \cos^2 \pi g_+ \sin^2 \pi k   
\end{equation}
where $g_\pm =\frac{1}{2}(g\pm g')$.
Clearly this reduces to the previous result (\ref{ggspectrum})
as $g' \to g$ but when $g\ne g'$ there are important new features.  
In particular, there are additional gaps in the spectrum as shown in 
\fref{fig:genspectrum}.

\section{Bulk scattering amplitudes}
\label{bulkamplitudes}

In this section we consider scattering amplitudes involving bulk fields.
General bulk amplitudes are functions of the boundary coupling $g$ and
our goal is to determine this dependence.  Some bulk amplitudes that are
zero in the free theory are nonvanishing in the presence of the boundary
interaction.  The periodic boundary potential breaks translation invariance 
in the target space and can absorb momenta in $\sqrt{2}\,\mathbb{Z}$.

Callan \etal \cite{Callan:1994ub} gave a simple prescription for 
scattering amplitudes that describe elementary 
string excitations reflecting off the interacting worldsheet boundary.   
Their method, which can be called {\em the method of rotated images},
rests on the fact that the bulk operators $\partial \phi$ and 
$\bar\partial\phi$, which create and destroy left- and
right-moving excitations, are in fact currents of the underlying
$SU(2)$ algebra.  The method can be generalized to deal with bulk
scattering amplitudes involving operators, which carry well defined $SU(2)$ 
quantum numbers, {\it i.e.}\ the discrete bulk operators 
$\Psi_{j \bar\jmath m}$ in equation (\ref{discretebulkfield}).

Consider some number of these states inserted in the upper half-plane and 
use the method of images so that for each insertion
$\bar\psi_{\bar\jmath m}(\bar z) \to \psi_{\bar\jmath m}(z^*)$.
This is possible even with the boundary interaction
turned on because the anti-holomorphic field 
$\bar\psi_{\bar\jmath m}(\bar z)$ commutes with the holomorphic $SU(2)$ 
currents in the interaction and reflects through the underlying Neumann 
boundary condition exactly as in the free theory.
In a perturbative expansion of the bulk amplitude the $SU(2)$ currents 
that appear in the boundary interaction are repeatedly integrated along
the real axis.  Their integration contours may be deformed away from 
the real axis into the lower-half-plane, where they will act on the
image fields $\psi_{\bar\jmath m}(z^*)$. 
The $SU(2)$ current algebra ensures that there will be 
no cuts generated, and so the integration contours can be closed around 
the image fields.  The net effect is the global $SU(2)$ rotation $U(g)$, 
given at the end of section~\ref{currentalgebra}, acting on each image field insertion.
The original right-moving discrete field was a component of a rank 
$\bar\jmath$ irreducible tensor operator, so the rotated image is
$\widetilde \psi_{\bar\jmath m}(z^*)
= \sum_{m'=-\bar\jmath}^{\bar\jmath}
\mathcal{D}^{\bar\jmath}_{m,m'}(g) \psi_{\bar\jmath m'}(z^*)$
with the rotation coefficient given by
\begin{equation}
\mathcal{D}^{\bar\jmath}_{m,m'}(g)=
\langle \bar\jmath,m\vert U(g) \vert \bar\jmath,m'\rangle
= \langle \bar\jmath,m\vert
e^{\pi i (g J_+ + \bar g J_-)} \vert \bar\jmath,m'\rangle\,,
\end{equation}
where $\vert j,m\rangle$ are standard $SU(2)$ states.

A bulk $n$-point function is then expressed in terms of
$2n$ point functions of free holomorphic fields and $SU(2)$
rotation coefficients,
\begin{eqnarray}
\fl \langle \Psi_{j_1\bar\jmath_1 m_1}(z_1,\bar z_1)\ldots
\Psi_{j_n \bar\jmath_n m_n}(z_n,\bar z_n)\rangle \nonumber \\
= \sum_{m'_i=-\bar\jmath_i}^{\bar\jmath_i}
\mathcal{D}^{\bar\jmath_1}_{m_1,m'_1}(g)
\ldots \mathcal{D}^{\bar\jmath_n}_{m_n,m'_n}(g)\,
\langle \psi_{j_1 m_1}(z_1)\psi_{\bar\jmath_1 m'_1}(z_1^*)
\ldots\rangle\bigg\vert_{g=0} \,.
\end{eqnarray}

As a simple example of this prescription we consider the one-point
function of a general discrete bulk primary,
\begin{equation}
\langle \Psi_{j \bar\jmath m}(z,\bar z)\rangle =
\sum_{m'=-\bar\jmath}^{\bar\jmath} \mathcal{D}^{\bar\jmath}_{m,m'}(g)
\langle \psi_{j m}(z) \psi_{\bar\jmath m'}(z^*) \rangle\bigg\vert_{g=0}\,.
\end{equation}
Conformal invariance
requires the scaling dimension of the two chiral operators to be the
same, i.e. $\bar\jmath^2 = j^2$, and momentum conservation in the
free chiral theory requires $m'=-m$.  The one-point function is therefore
\begin{equation}
\langle \Psi_{j\bar\jmath m}(z,\bar z) \rangle
=\delta_{j,\bar\jmath}\,
\frac{\mathcal{D}^{j}_{\bar m,-m}(g)}{(z-z^*)^{2j^2}} \,.
\end{equation}
For comparison, note that in the free theory with Neumann boundary 
conditions the only bulk operator 
that has a non-vanishing one-point function is the unit operator.  

Higher-point bulk amplitudes involving discrete fields can be computed 
in an analogous fashion, but unfortunately our prescription can not be 
applied to all bulk amplitudes in the model.  In general, bulk amplitudes 
will involve both the discrete bulk fields and fields carrying generic 
target space momenta.  The only constraint from momentum conservation is 
that the total momentum of all the fields in a given correlator add up to
an integer times $\sqrt{2}$ in our units.  The operator product of the 
currents in the boundary interaction and generic momentum fields is 
non-local, and so the effect of the interaction is no longer captured by 
a global $SU(2)$ rotation.

\section{Boundary amplitudes}
\label{boundarycorrelators}

The physics of the open string sector is encoded in correlation functions
of boundary operators as discussed in section~\ref{boundaryfields}.
It is straightforward to identify the boundary operators in
the free theory.  Consider bringing a bulk primary operator at generic 
momentum $\Psi(z,\bar z)=\exp (\rmi p [\phi(z)+\bar\phi(\bar z)])$ 
close to the boundary.  Now replace the right-moving part by its 
left-moving image and apply the left-moving OPE,
$\exp(\rmi p\phi(z))\exp(\rmi p\phi(z^*)) 
= (z-z^*)^{p^2}\exp(\rmi p\Phi(x)) + \ldots$,
where $x = \half (z+z^*)$ and the boundary scalar field is related to the
left-moving field by $\Phi(x)=2\phi(z)\vert_{z=z*}$.
For the discrete bulk fields one obtains
\begin{equation}
\fl\Psi_{j\bar\jmath m}(z,\bar z) \to \psi_{jm}(z)\psi_{\bar\jmath m}(z^*)
=\sum_{J=|j-\bar\jmath|}^{j+\bar\jmath}
\frac{A^{JM}_{jm;\bar\jmath m}}{(z-z^*)^{j^2+\bar\jmath^2-J^2}}\Psi^0_{JM}(x)
+ \ldots
\label{eq:dbb}
\end{equation}
The $\Psi^0_{JM}(x)$ are primary fields on the boundary, which we 
refer to as {\em discrete boundary fields}.  The superscript on
$\Psi^0_{JM}$ signals that these are boundary operators of the free theory.
The bulk-to-boundary OPE coefficients $A^{JM}_{jm;\bar\jmath m}$ can be
obtained by straightforward calculation.  By momentum conservation they
vanish unless $M=2m$.
The discrete boundary fields inherit the $SU(2)$ structure from the chiral
discrete fields, but since $j-\bar\jmath \in \mathbb{Z}$ we find that only
integer values of $J$ are allowed on the boundary.  The boundary scaling 
dimension of $\Psi^0_{JM}$ is $J^2$.

We are interested in calculating amplitudes involving arbitrary boundary
fields in the interacting theory, including boundary condition changing fields.
This is delicate since operators are now inserted on the boundary
where the non-linear self-interaction takes place.  We can nevertheless 
anticipate the structure of low-order amplitudes from conformal 
symmetry.  For two- and three-point functions one finds
\begin{eqnarray}
\langle \Psi_i(x_1) \Psi_j(x_2)\rangle 
= \frac{G_{ij}}{|x_1-x_2|^{\Delta_i + \Delta_j}} , \\
\fl \langle \Psi_i(x_1) \Psi_j(x_2)\Psi_k(x_3)\rangle 
= \frac{C_{ijk}}{
|x_1-x_2|^{\Delta_i + \Delta_j - \Delta_k}
|x_2-x_3|^{\Delta_j + \Delta_k - \Delta_i}
|x_3-x_1|^{\Delta_k + \Delta_i - \Delta_j}} ,\nonumber
\end{eqnarray}
where $\Delta_i(g)$ is the scaling dimension obtained from equation
\eref{eq:generalspectrum}. It remains to determine the 
$g$-dependence of the coefficients $G_{ij}$ and $C_{ijk}$.

Explicit calculations for fields carrying arbitrary momenta are 
difficult because such fields are non-local with respect to the
$SU(2)$ currents in the boundary interaction, but we can proceed
further with amplitudes involving the discrete boundary fields.
A key observation, that can be read off from equation \eref{ggspectrum},
is that the boundary scaling dimension of $\Psi^g_{JM}$ is $J^2$,
independent of the coupling. As a result we can write $\Psi^g_{JM}$
as a linear-combination of free boundary operators within the same
$SU(2)$ multiplet
\begin{equation}
\Psi^g_{JM}(x) = \sum_{M' = -J}^{J}h^{J}_{MM'}(g)\Psi^0_{JM'}(x)\,.
\label{eq:psig}
\end{equation}
We now write the general 
bulk-to-boundary OPE at non-zero boundary coupling,
\begin{equation}
\Psi_{j\bar\jmath m}(z,\bar z) 
= \sum_{J=|j-\bar\jmath |}^{j+\bar\jmath} \sum_{M=-J}^{J} 
\frac{A^{JM}_{jm;\bar\jmath m}(g)}{(z-z^*)^{j^2+\bar\jmath^2-J^2}}
\Psi^g_{JM}(x) + \ldots \,.
\label{eq:genbb}
\end{equation}
We then apply the method of rotated images as described in 
section~\ref{bulkamplitudes} on the left-hand side, plug 
in the expansion (\ref{eq:psig}) on the right-hand side, and
use equation (\ref{eq:dbb}). 
This results in a set of algebraic equations that 
relate the OPE coefficients $A^{JM}_{jm;\bar\jmath m}(g)$ and 
the expansion coefficients $h^J_{MM'}(g)$.  Unfortunately, there are not
enough equations to determine all the coefficients, so 
we need further input.

We propose the following prescription for determining $h^J_{MM'}(g)$.
First the free boundary operators $\Psi^0_{JM}$ are obtained as in equation 
\eref{eq:dbb}.  The effect of the interaction on these operators is 
then computed by letting the integration contours of the boundary currents
approach the real axis, where $\Psi^0_{JM}$ is inserted, from above.
The contours are then moved into the lower-half-plane, resulting in an $SU(2)$
rotation acting on $\Psi^0_{JM}$, giving 
$h^J_{MM'}(g) = \mathcal{D}^J_{MM'}(g)$.
This leads to boundary amplitudes of the form 
\begin{eqnarray}
\label{boundaryamplitudes}
\fl\langle\Psi^g_{J_1M_1}(x_1)\ldots\Psi^g_{J_nM_n}(x_n)\rangle\nonumber\\
=\sum_{M'_i=-J_i}^{J_i}
\mathcal{D}^{J_1}_{M_1M'_1}(g)\ldots\mathcal{D}^{J_n}_{M_nM'_n}(g)
\langle\Psi^0_{J_1M'_1}(x_1)\ldots\Psi^0_{J_nM'_n}(x_n)\rangle
\end{eqnarray}
and bulk-to-boundary OPE coefficients
$A^{JM}_{jm;\bar\jmath m}(g) 
= \mathcal{D}^j_{m,M-m}(-g)A^{JM}_{jm;\bar\jmath,M-m}(0) \,.$
We stress that other prescriptions are possible.  One alternative would be 
to deform the integration contours into the upper-half-plane in which case 
$\Psi^g_{JM} = \Psi^0_{JM}$ and the effect of the interaction is shifted 
entirely into the bulk-to-boundary coefficients $A^{JM}_{jm;\bar\jmath m}(g)$. 
We note, however, that the boundary amplitudes (\ref{boundaryamplitudes}) 
have the desirable feature that momentum is only conserved modulo $\sqrt{2}$ 
which reflects broken translation invariance.  

\ack
We thank C.~Callan, D.~Friedan and I.~Klebanov for discussions.
This work was supported in part by grants from the Icelandic Research 
Council, The University of Iceland Research Fund and The Icelandic 
Research Fund for Graduate Students.  L.T.~thanks the New High Energy 
Theory Center at Rutgers University for hospitality.

\end{document}